\documentclass[12pt]{article}
\usepackage{epsfig}
\usepackage{color}
\usepackage{lineno}

\begin{document}

\begin{center}

\vspace*{1.0cm}

{\Large \bf{First search for $\alpha$ decays of naturally
occurring Hf nuclides with emission of $\gamma$ quanta}}

\vskip 1.0cm

{\bf F.A.~Danevich$^{a,}$\footnote{Corresponding author. {\it
E-mail address:} danevich@kinr.kiev.ua (F.A.~Danevich).},
M.~Hult$^{b}$, D.V.~Kasperovych$^{a}$, G.P.~Kovtun$^{c,d}$,
K.V.~Kovtun$^{e}$, G.~Lutter$^{b}$, G.~Marissens$^{b}$,
O.G.~Polischuk$^{a}$, S.P.~Stetsenko$^{c}$, V.I.~Tretyak$^{a}$}

\vskip 0.3cm

$^{a}${\it Institute for Nuclear Research, 03028 Kyiv, Ukraine}

$^{b}${\it European Commission, Joint Research Centre, Retieseweg
111, 2440 Geel, Belgium}

$^{c}${\it National Scientific Center Kharkiv Institute of Physics
and Technology, 61108 Kharkiv, Ukraine}

$^{d}${\it Karazin Kharkiv  National University, 61022 Kharkiv,
Ukraine}

$^{e}${\it Public Enterprise ``Scientific and Technological Center
Beryllium'', 61108 Kharkiv, Ukraine}

\end{center}


\vskip 0.5cm

\begin{abstract}

The first ever search for $\alpha$ decays to the first excited
state in Yb was performed for six isotopes of hafnium (174, 176,
177, 178, 179, 180) using a high purity Hf-sample of natural
isotopic abundance with a mass of 179.8 g. For $^{179}$Hf, also
$\alpha$ decay to the ground state of $^{175}$Yb was searched for
thanks to the $\beta$-instability of the daughter nuclide
$^{175}$Yb. The measurements were conducted using an ultra
low-background HPGe-detector system located 225 m underground.
After 75 d of data taking no decays were detected but lower bounds
for the half-lives of the decays were derived on the level of
$\lim T_{1/2}\sim 10^{15}-10^{18}$~a. The decay with the shortest
half-life based on theoretical calculation is the decay of
$^{174}$Hf to the first $2^+$ 84.3~keV excited level of
$^{170}$Yb. The experimental lower bound was found to be
$T_{1/2}\geq 3.3\times 10^{15}$ a.

\end{abstract}

\vskip 0.4cm


\section{INTRODUCTION}

Alpha decay is one of the most important topics of nuclear physics
both from the theoretical and experimental points of view. The
process played a crucial role in the development of nuclear
models, since it offers information about the nuclear structure,
the nuclear levels and the properties of nuclei. In the last two
decades many experimental and theoretical studies have been
performed to investigate very rare $\alpha$ decays with long
half-lives ($10^{15}-10^{20}$ a) or/and low branching ratios
(10$^{-3}-10^{-8}$) (see review \cite{Belli:2019} and references
therein). The progress has been become possible thanks to the
substantial improvements in the experimental low-background
techniques and underground location of the experimental set-ups.

Natural hafnium consists of 6 isotopes, all of them are
theoretically unstable in relation to $\alpha$ decay; the energy
releases ($Q_{\alpha}$) are in the range of $1.3-2.5$~MeV (see
Table~1). The $\alpha$ decay to the ground state (g.s.) of the
daughter nuclide was experimentally observed only for $^{174}$Hf
which has the biggest value of $Q_{\alpha}=2494.5$ keV. Riezler
and Kauw detected the decay in 1959 with the half-life
$T_{1/2}=4.3\times10^{15}$ a by using nuclear emulsions
\cite{Riezler:1959}. Then Macfarlane and Kohman measured the
half-life as $T_{1/2}=2.0(4)\times10^{15}$ a with the help of an
ionization chamber \cite{Macfarlane:1961}. It should be noted that
both the experiments utilized samples of enriched $^{174}$Hf
(10.14\% in both cases). The exposure in the experiments for the
isotope $^{174}$Hf was $0.0068$ g$\times$d \cite{Riezler:1959} and
0.12 g$\times$d \cite{Macfarlane:1961}, respectively. The
half-life value obtained in \cite{Macfarlane:1961} is accepted
currently as the recommended half-life of $^{174}$Hf
\cite{Baglin:2018}.

\begin{table}[ht]
\caption{Characteristics of (potential) $\alpha$ decays of hafnium
isotopes. $\delta$ is the isotopic abundance of the nuclide in the
natural isotopic composition of elements. $Q_{\alpha}$ value is
given for the g.s. to g.s. transitions. Number of nuclei of the
isotope of interest in the hafnium sample used in the present
study is denoted as $N$.}
\begin{center}
\begin{tabular}{|c|c|c|c|}
\hline
Transition                  & $\delta$ \cite{Meija:2016} & $Q_{\alpha}$ (keV) \cite{Wang:2017} & $N$ \\
\hline
$^{174}$Hf $\to$ $^{170}$Yb & 0.0016(12)        & 2494.5(23)         & $9.71\times10^{20}$ \\
$^{176}$Hf $\to$ $^{172}$Yb & 0.0526(70)        & 2254.2(15)         & $3.19\times10^{22}$ \\
$^{177}$Hf $\to$ $^{173}$Yb & 0.1860(16)        & 2245.7(14)         & $1.13\times10^{23}$ \\
$^{178}$Hf $\to$ $^{174}$Yb & 0.2728(28)        & 2084.4(14)         & $1.65\times10^{23}$ \\
$^{179}$Hf $\to$ $^{175}$Yb & 0.1362(11)        & 1807.7(14)         & $8.26\times10^{22}$ \\
$^{180}$Hf $\to$ $^{176}$Yb & 0.3508(33)        & 1287.1(14)         & $2.13\times10^{23}$ \\
\hline
\end{tabular}
\label{isotopes}
\end{center}
\end{table}

\begin{figure}[ht]
\begin{center}
 \mbox{\epsfig{figure=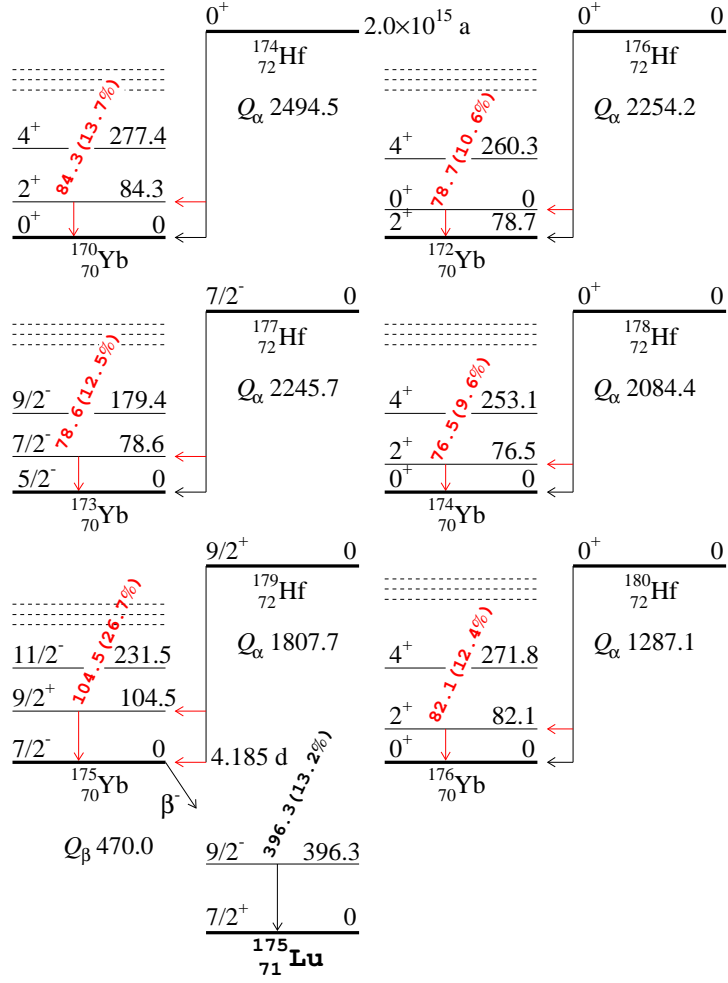,height=13.0cm}}
 \caption{(Color online) Expected schemes of $\alpha$ decay of the naturally
occurring hafnium isotopes (levels above the second excited levels
are omitted). The $Q_\alpha$ values, energies of the levels and of the de-excitation
$\gamma$ quanta are given in keV, the probabilities of $\gamma$
quanta emission are given in parentheses.
Red arrows show transitions investigated in this work.}
 \label{fig:dec-shem}
\end{center}
\end{figure}

In addition to $\alpha$ decay to the ground state, all the
isotopes can decay to excited levels of the daughter nuclei. In
the latter processes the deexcitation $\gamma$ quanta are emitted,
which can be searched for by low-background $\gamma$-ray
spectrometry. Because of the exponential dependence of the
half-life on the decay energy [according to the Geiger-Nuttall law
$T_{1/2}\sim exp(Q_{\alpha}^{-1/2})$] it is experimentally
difficult to search for decay branches with low
$Q_{\alpha}$-value. Therefore, in this work, we look for
transitions only to the first excited levels of the daughter Yb
nuclides. However, in case of the decay $^{179}$Hf $\to$
$^{175}$Yb, the $^{175}$Yb daughter is unstable. It beta decays to
$^{175}$Lu ($T_{1/2} \approx 4.2$ d and $Q_{\beta}= 470$ keV) with
the emission of gamma quanta, being the most intensive one with
$E_{\gamma}=396.3$ keV \cite{Basunia:2004}. Thus, also $^{179}$Hf
decay to the ground state of $^{175}$Yb can be looked for in such
an approach. The simplified decay schemes of Hf $\alpha$ decays
are shown in Fig. \ref{fig:dec-shem}.

\nopagebreak

\pagebreak

In this work, a search for $\alpha$ decays accompanied by
gamma-ray emission of six naturally occurring Hf isotopes was
conducted. A specially purified hafnium sample (179.8 g) of
natural isotopic abundance was measured in a HPGe-detector system
located 225 m underground in the laboratory HADES (Belgium). The
obtained $T_{1/2}$ limits are compared with theoretical
predictions based on a few theoretical models
\cite{Poenaru:1983,Buck:1991,Buck:1992,Denisov:2015}.

\section{EXPERIMENT}

\subsection{Sample of hafnium}

A disc-shaped sample of metallic hafnium with diameter of 59.0 mm
and 5.0 mm height, with the mass of 179.8 g, was used in the
low-background experiment. The hafnium was obtained by reduction
process from hafnium tetrafluoride with metallic calcium. Then the
material was additionally purified by double melting in vacuum by
electron beam.

The purity of the obtained hafnium was measured by the Laser
Ablation Mass Spectrometry as $\simeq99.8\%$. It should be
stressed that zirconium is typically the main contaminant of
hafnium (in our case the mass concentration of Zr is 0.4\%).
However, the purity level of Hf is usually given without taking
into account the Zr contamination. The concentrations (limits) of
other metal and gaseous impurities in the sample were on the level
of $0.005-0.05\%$ ($\approx0.2\%$ in total). The summary of the
impurities detected (or their limits) in the Hf sample is
presented in Table \ref{tab:cont}.

\nopagebreak
\begin{table}[htb]
\caption{Impurities detected (bounded) in the Hf sample by the
Laser Ablation Mass Spectrometry.}
\begin{center}
\begin{tabular}{|c|c|}

 \hline
 Element    & Mass concentration (\%) \\
 \hline
 C          & $<0.01$ \\
 N          & $<0.005$ \\
 O          & $<0.05$ \\
 Mg         & $<0.004$ \\
 Al         & $<0.005$ \\
 Si         & $<0.005$ \\
 Ca         & 0.01 \\
 Ti         & $<0.005$ \\
 Cr         & $<0.003$ \\
 Mn         & $<0.0005$ \\
 Fe         & 0.04 \\
 Ni         & 0.02 \\
 Cu         & $<0.005$ \\
 Zr         & 0.4 \\
 Nb         & 0.01 \\
 Mo         & 0.01 \\
 W          & 0.01 \\
  \hline

\hline
\end{tabular}
\label{tab:cont}
\end{center}
\end{table}

\subsection{Gamma-ray spectrometry set-ups}

\begin{figure}[ht]
\begin{center}
 \mbox{\epsfig{figure=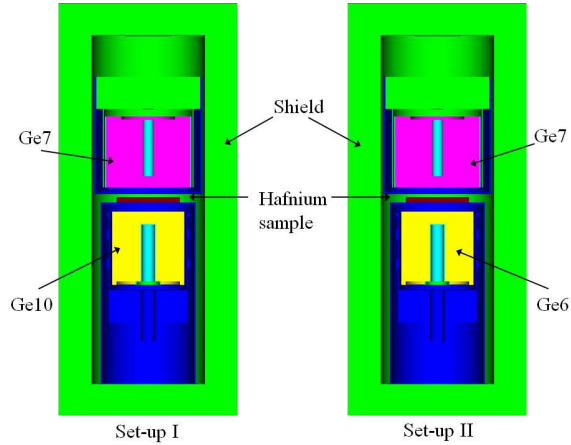,height=6.0cm}}
\caption{(Color online) Schematic view of the two low-background
set-ups with HPGe detectors and hafnium sample.}
 \label{fig:set-ups}
\end{center}
\end{figure}

The experiment was realized with the help of two set-ups with
three HPGe detectors (named Ge6, Ge7, and Ge10) at the HADES
underground laboratory (Geel, Belgium) located at depth of 225 m
below the ground. A schematic view of the set-ups is given in Fig.
\ref{fig:set-ups}. The main characteristics of the detectors are
presented in Table \ref{tab:detectors} (more details one can find
in Refs. \cite{Wieslander:2009,Hult:2013}). First the Hf sample
was stored 13 d underground to enable decay of short-lived
cosmogenic radionuclides. The first measurement in which the
sample was installed directly on the endcap of the detector Ge10
lasted 40.4 d (the set-up I, see Fig. \ref{fig:set-ups}). The Ge10
detector is perfect for looking for low-energy $\gamma$-rays (the
effects searched for) taking into account its excellent energy
resolution and high detection efficiency to low-energy $\gamma$
quanta. Then the experiment was continued with the Ge6 detector
instead of Ge10 for 34.8 d (set-up II). The Ge6 detector has a
higher detection efficiency to middle-energy and high-energy
$\gamma$ quanta thanks to a bigger volume and was used in the 2nd
stage of the experiment to investigate in wider energy range
radioactive contamination of the Hf sample thoroughly. The endcap
height of the Ge10 detector is lower than that of Ge6. Therefore
Ge7 was slightly (1.1 mm) further away from the sample in the
set-up I compared to the set-up II. This consequently resulted in
a higher detection efficiency of the detector Ge7 in the set-up II
(see Table \ref{tab:res174} in Section \ref{sec:res174Hf}). The
total exposure of the experiment is 42 g$\times$d for the isotope
$^{174}$Hf (the exposure was calculated as a product of the
$^{174}$Hf isotope mass in the sample and of the sum of the four
measuring times of the detectors).

\begin{table}[htb]
\caption{Properties of the HPGe detectors used in this study.}
\begin{center}
\begin{tabular}{|l|c|c|c|}
 \hline
 ~                                      & Ge6           & Ge7       & Ge10 \\
 \hline

 Energy resolution (FWHM) at 84 keV     & 1.4 keV       & 1.3 keV   & 0.9 keV \\

FWHM at 396 keV                         & 1.8 keV       & 1.5 keV   & 1.2 keV \\

FWHM at 1332 keV                        & 2.3 keV       & 2.2 keV   & 1.9 keV \\

 Relative efficiency                    & 80\%          & 90\%      & 62\%  \\

 Crystal mass                           & 2096 g        & 1778 g    & 1040 g  \\

 Window material and thickness          & LB Cu 1.0 mm & HPAl 1.5 mm & HPAl 1.5 mm \\

 Top dead layer thickness               & 0.9 mm        & 0.3 $\mu$m    & 0.3 $\mu$m  \\

\hline
\multicolumn{4}{l}{LB Cu = Low Background Copper} \\
\multicolumn{4}{l}{HPAl = High Purity Aluminum} \\

\end{tabular}
\label{tab:detectors}
\end{center}
\end{table}

\section{RESULTS AND DISCUSSION}

\subsection{Radioactive impurities in the hafnium sample}
\label{sec:rad-cont}

\begin{figure}[htb]
\begin{center}
 \mbox{\epsfig{figure=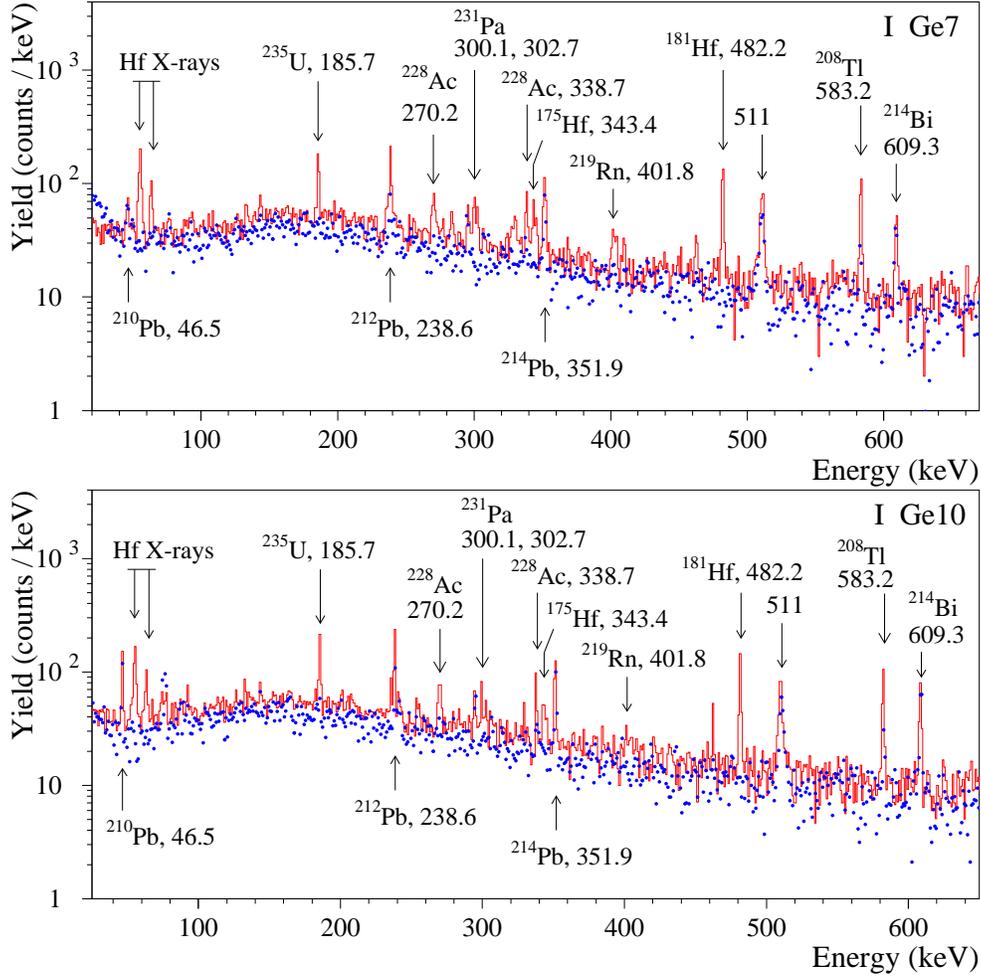,height=13.0cm}}
 \caption{(Color online) Energy spectra accumulated with the hafnium
sample (solid line) and without sample (dots) by
ultra-low-background HPGe $\gamma$ detectors Ge7 (over 38.4 d with
the hafnium sample and over 38.5 d without sample), and Ge10 (over
40.4 d with hafnium and 38.5 d background). The background energy
spectra are normalized to the times of measurements with the Hf
sample. Energy of $\gamma$ lines are in keV.}
 \label{fig:set-up1}
\end{center}
\end{figure}

\begin{figure}[ht]
\begin{center}
 \mbox{\epsfig{figure=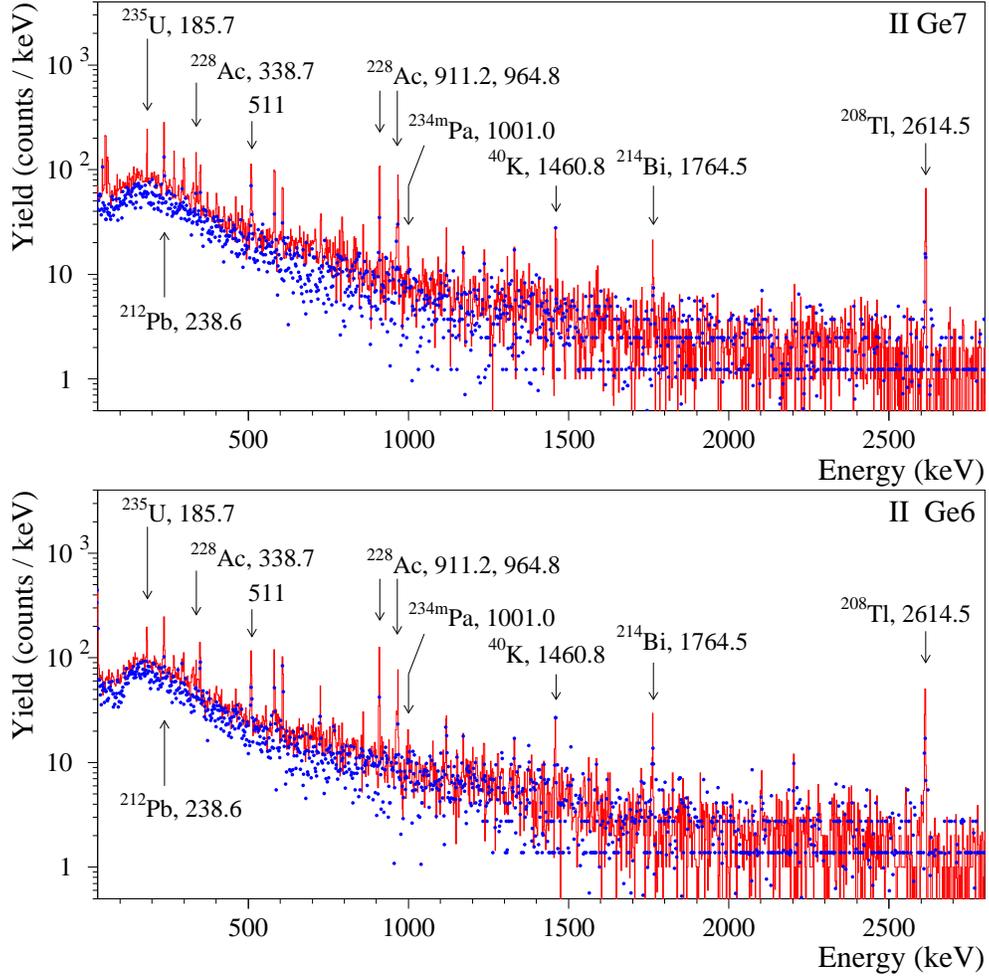,height=13.0cm}}
 \caption{(Color online) Energy spectra accumulated with the hafnium
sample (solid line) and without sample (dots) by
ultra-low-background HPGe $\gamma$ detectors Ge7 (over 34.8 d with
the hafnium sample and over 28.1 d without sample), and Ge6 (over
34.8 d with hafnium and 25.3 d background). The background energy
spectra are normalized to the times of measurements with the Hf
sample. Energy of $\gamma$ lines are in keV.}
 \label{fig:set-up2}
\end{center}
\end{figure}

Energy spectra recorded by the HPGe detectors with the hafnium
sample and without sample (background) are presented in
Fig.~\ref{fig:set-up1} (set-up I) and Fig.~\ref{fig:set-up2}
(set-up II). The gamma peaks identified in the spectra belong
mainly to the naturally occurring primordial radionuclides:
$^{40}$K, and daughters of the $^{232}$Th, $^{235}$U, and
$^{238}$U families. There is statistically significant excess in
the peaks of $^{228}$Ac (in equilibrium with $^{228}$Ra from the
$^{232}$Th family), $^{235}$U, $^{231}$Pa, $^{227}$Ac (daughters
of $^{235}$U), $^{234m}$Pa and $^{226}$Ra ($^{238}$U) in the data
collected with the sample. In addition, we have observed in the
hafnium sample two cosmogenic (neutron induced) radionuclides
$^{175}$Hf [decays by electron capture with $Q_{EC}=683.9(20)$ keV
and half-life $T_{1/2}=70(2)$ d] and $^{181}$Hf [beta active with
$Q_{\beta}=1035.5(18)$ keV, $T_{1/2}=42.39(6)$ d]. It should be
stressed that the counting rate in the $\gamma$ peaks of
$^{175}$Hf and $^{181}$Hf became substantially lower in the later
measurements in the set-up II due to decay in the underground
conditions.

\clearpage

Massic activities of the radionuclides in the hafnium sample were
calculated with the formula:

\begin{equation}
A = (S_{sample}/t_{sample}-S_{bg}/t_{bg})/(\eta \cdot \varepsilon
\cdot m)
 \end{equation}

\noindent where $S_{sample}$ ($S_{bg}$) is the area of a peak in
the sample (background) spectrum; $t_{sample}$ ($t_{bg}$) is the
time of the sample (background) measurement; $\eta$ is the
$\gamma$-ray emission intensity of the corresponding transition
\cite{Firestone:1996}; $\varepsilon$ is the full energy peak
efficiency; $m$ is the mass of the sample. The detection
efficiencies were calculated with EGSnrc simulation package
\cite{Kawrakow:2017,Lutter:2018}, the events were generated
homogeneously in the Hf sample. The calculations were validated in
the measurements with $^{109}$Cd and $^{133}$Ba $\gamma$ sources
(set-up I), and $^{109}$Cd, $^{133}$Ba, $^{134}$Cs, $^{152}$Eu,
and $^{241}$Am sources (set-up II). The standard deviation of the
relative difference between the simulations and the experimental
data is $5-7\%$ for $\gamma$ peaks in the energy interval 53.2 keV
-- 383.8 keV for the set-up I, and is 6\% for $\gamma$ peaks in
the energy interval 59.5 keV -- 1408.0 keV for the set-up II.

A clear excess of the count-rate in the energy spectra gathered
with the Hf sample observed for $\gamma$ peaks of $^{228}$Ac (in
equilibrium with $^{228}$Ra), $^{212}$Pb, $^{212}$Bi and
$^{208}$Tl ($^{228}$Th) allows to estimate the massic activities
in the sample of $^{228}$Ra ($20\pm4$ mBq/kg) and $^{228}$Th
($13.3\pm1.4$ mBq/kg). No wonder that the equilibrium of the
$^{232}$Th chain is broken: the chemical properties of radium and
thorium are rather different.

A massic activity of $^{238}$U was estimated on the basis of
$^{234m}$Pa $\gamma$ peak with energy 1001.0 keV as $59\pm28$
mBq/kg. Some excess of $^{214}$Pb and $^{214}$Bi $\gamma$-peaks
areas was observed in the energy spectra accumulated with the Hf
sample. Assuming the $^{226}$Ra sub-chain in equilibrium,
$^{226}$Ra massic activity can be calculated as $1.6\pm1.3$
mBq/kg. It should be stressed, however, that the measurements were
not carried out in a radon-tight container. Thus, we give
conservatively a limit on the $^{226}$Ra massic activity in the
sample: $\leq3.7$ mBq/kg. The substantially smaller activity of
$^{226}$Ra in comparison to its mother $^{238}$U can be explained
by a broken equilibrium of the $^{238}$U chain occurred in the
hafnium sample production process, also a quite expectable case
taking into account the rather different chemical properties of
uranium and radium. We cannot conclude that the $^{210}$Pb
sub-chain is out of equilibrium too. However, the sensitivity of
the set-ups to $^{210}$Pb is rather limited due to the low
detection efficiency and small $\gamma$-ray emission intensity of
the 46.5 keV gamma quanta of $^{210}$Pb (4.25\%).

A significant deviation of the $^{235}$U/$^{238}$U activities
ratio [the ratio is $0.36(18)$] from the expected one (0.046,
assuming the natural isotopic abundance of the uranium isotopes)
was observed in the Hf sample. An explanation for this could be
that the production of the high pure hafnium tetrafluoride
included centrifugation of gaseous Hf compound to reduce zirconium
concentration. The contamination by $^{235}$U could occur due to
proximity between the industrial sites of the centrifugation
facilities to purify hafnium and to enrich uranium by the Soviet
Union industry. The details of the Hf production process are
unknown, however, this explanation of the $^{235}$U/$^{238}$U
ratio deviation looks a most reasonable one.

If no statistically significant peak excess was detected (the
cases of $^{40}$K, $^{60}$Co, $^{137}$Cs, $^{178m2}$Hf,
$^{182}$Hf), we set limits on massic activities of possible
radioactive impurities in the sample. The calculated massic
activities (limits) of radioactive impurities in the Hf sample are
summarized in Table \ref{tab:rad-cnt}. Only one unidentified peak
with energy 1310.6(4) keV and area 14(4) counts was observed in
the energy spectrum of the Ge7 detector measured with the Hf
sample in the set-up~II.

\nopagebreak
\begin{table}[h]
\caption{Radioactive contamination of the Hf sample measured in
HADES by HPGe $\gamma$-ray spectrometry. The massic activities of
$^{175}$Hf and $^{181}$Hf are given with reference date at the
start of each measurement for the set-up I and the set-up II
(within brackets) separately. The upper limits are given at 90\%
C.L., the reported uncertainties are the combined standard
uncertainties.}
\begin{center}
\begin{tabular}{|l|l|l|l|}

 \hline
  Chain     & Nuclide       & Massic activity (mBq/kg)                      &  Activity in the sample (mBq) \\
 \hline
 ~          & $^{40}$K      & $\leq 8$                                      & $\leq 1.4$ \\
 ~          & $^{60}$Co     & $\leq 0.6$                                    & $\leq 0.11$ \\
 ~          & $^{137}$Cs    & $\leq 1.1$                                    & $\leq 0.20$ \\
 ~          & $^{172}$Hf    & $\leq 17$                                     & $\leq 3$ \\
 ~          & $^{175}$Hf    & $2.5\pm0.3$ $(1.0\pm0.2)$                     & $0.44\pm0.05$ $(0.18\pm0.04)$  \\
 ~          & $^{178m2}$Hf  & $\leq 0.4$                                    & $\leq 0.06$ \\
 ~          & $^{181}$Hf    & $8.1\pm0.4$ $(0.6\pm0.2)$                     & $1.45\pm0.07$ $(0.12\pm0.04)$ \\
 ~          & $^{182}$Hf    & $\leq 2.8$                                    & $\leq 0.5$ \\
 \hline
 $^{232}$Th & $^{228}$Ra    & $20\pm4$                                      & $3.6\pm0.7$ \\
 ~          & $^{228}$Th    & $13.3\pm1.4$                                  & $2.38\pm0.25$ \\
 \hline
 $^{235}$U  & $^{235}$U     & $21\pm3$                                      & $3.8\pm0.5$  \\
 ~          & $^{231}$Pa    & $60\pm19$                                     & $11\pm3$  \\
 ~          & $^{227}$Ac    & $11\pm3$                                      & $2.0\pm0.5$ \\
 \hline
 $^{238}$U  & $^{234m}$Pa   & $59\pm28$                                     & $11\pm5$ \\
  ~         & $^{226}$Ra    & $\leq3.7$                                       & $\leq0.7$  \\
 ~          & $^{210}$Pb    & $\leq 280$                                    & $\leq50$ \\
  \hline
\hline
\end{tabular}
\label{tab:rad-cnt}
\end{center}
\end{table}

\subsection{Search for $\alpha$ decay of $^{174}$Hf to the first excited level of $^{170}$Yb}
\label{sec:res174Hf}

The energy release for $^{174}$Hf $Q_\alpha=2494.5$~keV is the
highest among the naturally occurring Hf nuclides, and its
$\alpha$ decay to the g.s. of  $^{170}$Yb was already observed
\cite{Riezler:1959,Macfarlane:1961}
($T_{1/2}=2.0\times10^{15}$~a). The energy of the first excited
level of $^{170}$Yb daughter is quite low (84.3~keV), and the
corresponding value of $Q_\alpha^*=2410.2$~keV is not far from
that for the g.s. transition that gives a hope to observe this
decay too. We pay a special attention to search for this process
here.

\begin{figure}[ht]
\begin{center}
 \mbox{\epsfig{figure=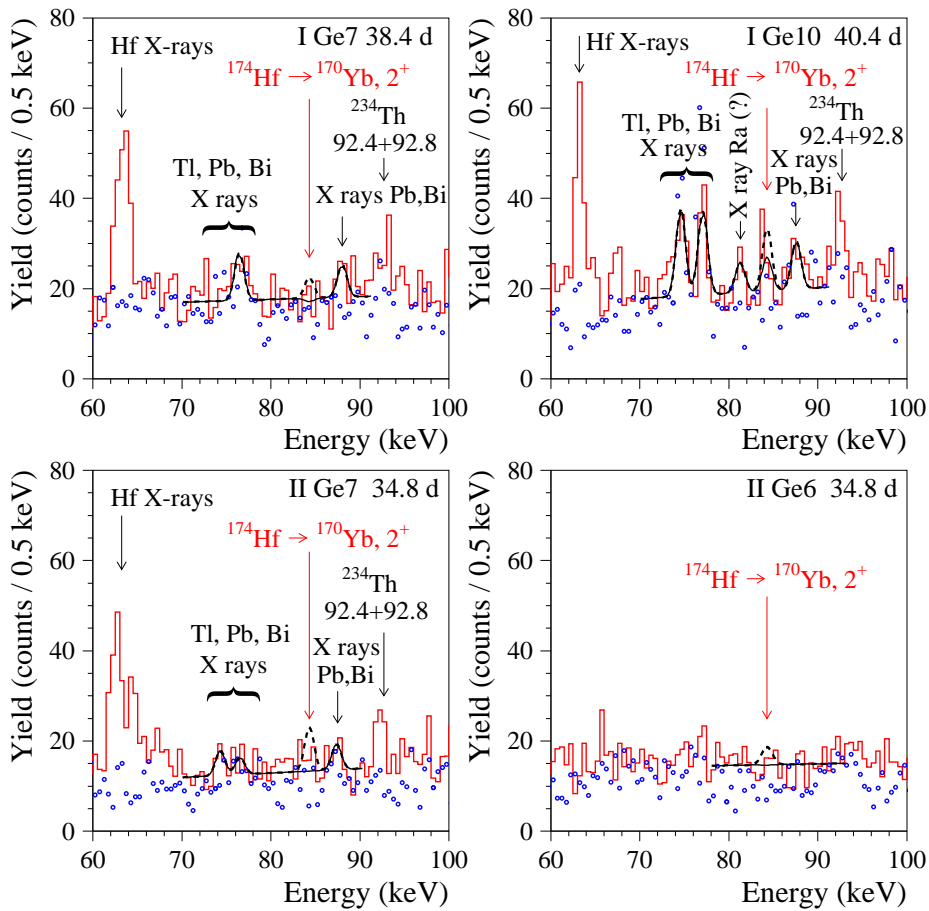,height=12.0cm}}
\caption{(Color online) Parts of the energy spectra accumulated
with the hafnium sample in the set-ups I and II by the HPGe
detectors (solid red histograms). The background energy spectra,
normalized on the time of measurements with the sample, are
depicted by blue dots. The models of the fits of the spectra with
sample are shown by solid lines, while the peaks expected in
$\alpha$ decay of $^{174}$Hf to the $2^+$ 84.3 keV excited level
of $^{170}$Yb, excluded with 90\% C.L., are shown by dashed
lines.}
 \label{fig:res-ind}
\end{center}
\end{figure}

There is no peak in any energy spectra that can be interpreted as
$\alpha$ decay of $^{174}$Hf to the 84.3 keV excited level of
$^{170}$Yb\footnote{In the spectrum of the detector Ge10 there is
a structure near the energy of interest with an area $20.2 \pm
10.2$ counts, that is no evidence of the effect searched for (see
text below and Fig. \ref{fig:res-ind}).}. Therefore we can set a
lower half-life limit on the decay with the following formula:

\begin{equation}
\lim T_{1/2} =  \ln 2 \cdot N \cdot \varepsilon \cdot \eta \cdot
t / \lim S,
\end{equation}

\noindent where $N$ is the number of potentially $\alpha$ unstable
nuclei, $\varepsilon$ is the detection efficiency, $\eta$ is the
84.3~keV gamma-ray emission intensity (equal to 13.7\%
\cite{Baglin:2018}), $t$ is the measurements time, and $\lim S$ is
the number of events of the effect searched for which can be
excluded at a given confidence level (C.L.; in the present work
all the half-life limits are given with 90\% C.L.). The detection
efficiencies for different detectors were simulated with the help
of the EGSnrc package \cite{Kawrakow:2017,Lutter:2018}.

To estimate the $\lim S$ value, the energy spectra accumulated
with the hafnium sample were fitted by a model consisting from the
effect searched for (a Gaussian peak centered at 84.3~keV with the
width determined for the each detector individually) and a first
order polynomial function to describe continuous background. To
extend the energy interval of fit (with an aim to improve
description of the background) we have added in the model X-ray
peaks. The X-rays can be caused by decays of $^{232}$Th, $^{235}$U
and $^{238}$U daughters in the sample and/or details of the
experimental set-ups. The experimental spectra were fitted in the
energy intervals within $70-80$~keV, for the starting point, and
$86-94$~keV, for the final point, with a step of 1 keV. The best
fits of the data obtained with the four detectors are shown in
Fig.~\ref{fig:res-ind}. The fits are characterized by the
$\chi^2/$n.d.f. values (where n.d.f. is the number of degrees of
freedom) within $0.54-0.93$. The obtained areas of the peak are
given in Table \ref{tab:res174} together with the values of $\lim
S$ (calculated by using recommendations \cite{Feldman:1998}), and
the detection efficiencies for the 84.3~keV gamma quanta in each
detector.

\begin{table}[h]
\caption{Half-life lower limits on $\alpha$ decay of $^{174}$Hf to
the 84.3 keV excited level of $^{170}$Yb obtained from analysis of
the data recorded with four HPGe detectors. The combined limit
obtained by using Eq.~(\ref{eq:lim-sum}) from the results of the
individual spectra fits, and from the fit of the sum spectrum of
the detector Ge7 in the set-ups I and II, and of the fit of sum
spectrum of all four detectors are presented too (see text for
details). All the limits are given at 90\% C.L.}
\begin{center}
\begin{tabular}{|l|l|l|l|l|l|}
 \hline
 Set-up                                     & Time of       & Number of counts  & $\lim S$  & Detection     & Half-life  \\
 Detector                                   & measurements  & in 84.3 keV peak  & (counts)  & efficiency    & limit (a) \\
 ~                                          & (d)           &  ~                &  ~        & ~             & ~ \\
  \hline
 I Ge7                                      & 38.4          & $-2.0\pm8.6$      & 12.2      & 0.005176      & $4.1\times10^{15}$ \\
 \hline
 I Ge10                                     & 40.4          & $20.2\pm10.2$     & 36.9      & 0.004491      & $1.2\times10^{15}$ \\
 \hline
 II Ge7                                     & 34.8          & $12.1\pm8.5$      & 26.0      & 0.005310      & $1.8\times10^{15}$ \\
 \hline
 II Ge6                                     & 34.8          & $-0.3\pm7.5$      & 12.0      & 0.001251      & $9.1\times10^{14}$ \\
 \hline
 Combined                                   & ~             & ~                 & ~         & ~             &  ~ \\
 all four detectors                         & 148.4         & $30.0\pm17.5$     & 58.7      & ~             & $2.6\times10^{15}$ \\
  \hline
 Combined                                   & ~             & ~                 & ~         & ~             & ~  \\
 I Ge7 + II Ge7                             & 73.2          & $10.1\pm12.1$     & 29.9      & ~             & $3.2\times10^{15}$ \\
\hline
 Fit of sum spectrum                        & ~             & ~                 & ~         & ~             &  ~ \\
 of all four detectors                      & 148.4         & $29.8\pm17.4$     & 58.3      & ~             & $2.6\times10^{15}$ \\
 \hline
 Fit of sum spectrum                        & ~             & ~                 & ~         & ~             &  ~ \\
 I Ge7 + II Ge7                             & 73.2          & $9.8\pm11.5$      & 28.7      & ~             & $3.3\times10^{15}$ \\
 \hline
\end{tabular}
\label{tab:res174}
\end{center}
\end{table}

\clearpage

As a next step, we tried to improve the experimental sensitivity
by combining the data obtained with the different detectors. One
can join results of several measurements by using the following
formula:

\begin{equation}
T_{1/2} = \ln 2 \cdot N \cdot \eta \cdot \Sigma (\varepsilon_i
\cdot t_i) / \Sigma S_i,
\end{equation}

\noindent where $\varepsilon_i$ are the detection efficiencies,
$t_i$ are the times of measurements for each detector, $\Sigma
S_i$ is the sum of events in the peak searched for in the data
accumulated with the detectors. Since the effect is not observed,
the formula is transformed to equation for a half-life limit and a
limit on number of events as following:

\begin{equation}
\lim T_{1/2} = \ln 2 \cdot N \cdot \eta \cdot \Sigma
(\varepsilon_i \cdot t_i) / \lim S,
 \label{eq:lim-sum}
\end{equation}

\noindent where $\lim S$ can be estimated by two approaches.
First, one can calculate an arithmetic sum of the numbers of
counts in the peak searched for, obtained from the fits of the
individual spectra. One could expect that a highest sensitivity
should be achieved by analysis of all the data of the detectors
Ge6, Ge7 and Ge10 in the both set-ups. However, the analysis does
not provide a highest sensitivity (see Table \ref{tab:res174}). It
can be explained by the larger background in the sum spectrum and
in average a lower detection efficiency (mainly of the detector
Ge6 that has rather thick dead layer and thick copper endcap). The
highest sensitivity was achieved by combining the peak areas
obtained with the detector Ge7 in the set-ups I and II,
$S=10.1\pm12.1$ counts, that results in the number of excluded
events $\lim S =29.9$ counts (according to the recommendations
\cite{Feldman:1998}) and the half-life limit
$T_{1/2}(^{174}$Hf$)\geq 3.2\times10^{15}$ a (at 90\% C.L.).

\begin{figure}[htb]
\begin{center}
 \mbox{\epsfig{figure=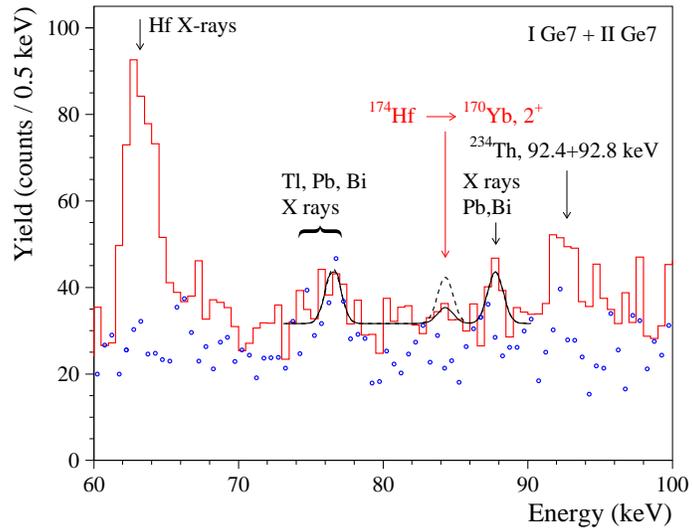,height=7.0cm}}
\caption{(Color online) Part of the sum energy spectrum
accumulated with the hafnium sample in the set-ups I and II by the
detector Ge7 over 73.2 d (solid red histogram), and the background
spectrum of the detector for 66.6 d (normalized on the time of
measurements with the sample, blue dots). Fit of the data with the
sample is shown by solid line, while the excluded with 90\% C.L.
peak expected in the $\alpha$ decay of $^{174}$Hf to the $2^+$
84.3 keV excited level of $^{170}$Yb with an area of 28.7 counts
is shown by dashed line.}
 \label{fig:res-sum}
\end{center}
\end{figure}

Another way to estimate the value of $\lim S$ is a fit of a sum
spectrum of several detectors, taking into account a rather
similar energy resolutions and the background conditions of the
detectors. Again, the highest sensitivity was obtained not by the
fit of all the sum spectra, but by the fit of the sum spectrum of
the detector Ge7 in the set-ups I and II (see Table
\ref{tab:res174}). The best fit was achieved in the energy
interval $73-90$ keV (with $\chi^2/$n.d.f. = 16.9/28 = 0.604),
providing the peak area $S=9.8\pm11.5$ counts that corresponds to
the $\lim S=28.7$ counts. The fit and the excluded effect are
shown in Fig.~\ref{fig:res-sum}. Finally, we accept the result of
the fit to derive the following limit on $\alpha$ decay of
$^{174}$Hf to the first excited level of $^{170}$Yb (at 90\%
C.L.):

\begin{center}
$T_{1/2}(^{174}$Hf$)\geq 3.3\times10^{15}$ a\footnote{One can see
in Table \ref{tab:res174} that the strongest half-life limit was
obtained with the detector Ge7 in the set-up I. However, the large
$\lim T_{1/2}$ appears due to the negative peak area that results
in a small $\lim S$ value. We prefer to use the estimation
obtained from the higher statistics acquired with the detector Ge7
in the set-ups I and II.}.
 \end{center}

\noindent The obtained limit is still about two-three orders of
magnitude lower than the theoretical estimations calculated by
using the methods proposed in
\cite{Poenaru:1983,Buck:1991,Buck:1992,Denisov:2015} (see Section
3.4).

\clearpage

\subsection{Search for $\alpha$ decay of other hafnium isotopes}

\begin{figure}[htb]
\begin{center}
 \mbox{\epsfig{figure=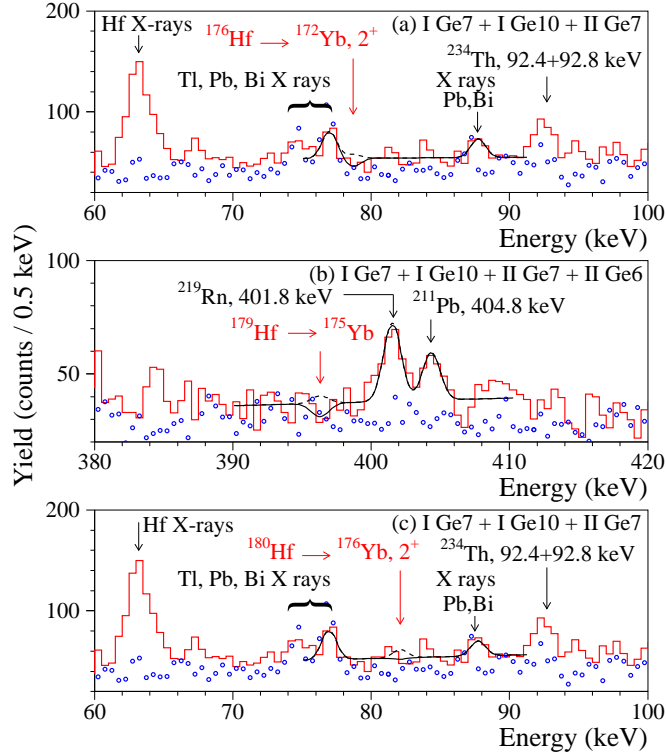,height=10.0cm}}
 \caption{(Color online) Parts of the sum energy spectra accumulated with the hafnium
sample by the detectors Ge7 (the set-ups I and II) and Ge10 (the
set-up I) with the total measurement time 113.6 d (a and c), and
of all four detectors over 148.4 d (b). The corresponding
background spectra, normalized on the time of measurements with
the Hf sample, are shown by blue dots. Fits of the data are shown
by solid lines, while the excluded with 90\% C.L. peaks expected
in the $\alpha$ decays of $^{176}$Hf (a), $^{179}$Hf (b) and
$^{180}$Hf (c) are drown by dashed lines.}
 \label{fig:res-other}
\end{center}
\end{figure}

To estimate half-life limits for other naturally occurring hafnium
nuclides relative to $\alpha$ decay to the first excited levels of
daughter nuclei, we have analyzed the individual spectra and
different combinations of the energy spectra in the energy
intervals where $\gamma$ peaks after the $\alpha$ decays are
expected. The data on the parameters of the fits, excluded peaks
areas and calculated half-life limits for $^{176}$Hf, $^{177}$Hf,
$^{178}$Hf, $^{179}$Hf and $^{180}$Hf are given in
Table~\ref{tab:other}. Examples of the sum energy spectra fits for
$^{176}$Hf, $^{179}$Hf and $^{180}$Hf are presented in
Fig.~\ref{fig:res-other}. As previously, the detection
efficiencies were calculated with the EGSnrc simulation package.
Gamma-ray emission intensities of $\gamma$'s expected in
deexcitation processes, are taken as: $^{176}$Hf --
$\eta$($\gamma$ 78.7 keV) = 10.6\% \cite{Singh:1995}; $^{177}$Hf
-- $\eta$($\gamma$ 78.6 keV) = 12.5\% \cite{Shirley:1995};
$^{178}$Hf -- $\eta$($\gamma$ 76.5~keV) = 9.6\%
\cite{Browne:1999}; $^{180}$Hf -- $\eta$($\gamma$ 82.1 keV) =
12.4\% \cite{Basunia:2006}.

In possible $\alpha$ decay of $^{179}$Hf, the $^{175}$Yb daughter
nucleus is unstable. It decays through $\beta^-$ decay to
$^{175}$Lu with $T_{1/2}=4.2$~d and $Q_\beta=470.0$~keV emitting
the most intensive $\gamma$ quantum with $E_\gamma=396.3$~keV with
yield of 13.2\% \cite{Basunia:2004}. Since the peak at 396.3~keV
is not detected in the measured energy spectra, we set the
following limit on $^{179}$Hf $\alpha$ decay half-life:
$T_{1/2}\geq2.2\times10^{18}$ a. It should be stressed that the
limit is valid also for $\alpha$ decay of $^{179}$Hf to the ground
state and for all the energetically allowed $\alpha$ transitions
to excited levels of $^{175}$Yb.

\begin{table}[ht]
\caption{Half-life lower limits on $\alpha$ decay of $^{176}$Hf,
$^{177}$Hf, $^{178}$Hf and $^{180}$Hf to the first excited levels
of the daughter Yb nuclei. Limit for $^{179}$Hf is valid for
$\alpha$ transitions to any of $^{175}$Yb levels, including the
ground state. The limits are given at 90\% C.L.}
\begin{center}
\begin{tabular}{|l|l|l|l|l|}
 \hline
 Transition                                 & Method                & Number of counts              & $\lim S$  &  Half-life  \\
                                            & and data              & in the expected $\gamma$ peak & (counts)  & limit (a) \\
                                            &                       & (the $\gamma$ peak energy     &           &  \\
                                            &                       & is given in parenthesis)      &           &  \\
 \hline
 $^{176}$Hf$~\rightarrow$~$^{172}$Yb        & Fit of sum spectra    & ~                             &  ~        & ~ \\
 $2^+$ 78.7 keV                             & I Ge7, I Ge10, II Ge7 & $-22.6\pm16.4$  (78.7 keV)    & 10.0      & $3.0\times10^{17}$ \\
 \hline
 $^{177}$Hf$~\rightarrow$~$^{173}$Yb        & Fit of sum spectra    & ~                             &  ~        & ~ \\
 $7/2^-$ 78.6 keV                           & I Ge7, I Ge10, II Ge7 & $-22.6\pm16.4$  (78.6 keV)    & 10.0      & $1.3\times10^{18}$ \\
 \hline
 $^{178}$Hf$~\rightarrow$~$^{174}$Yb        & Fit of sum of         & ~                             &  ~        & ~ \\
 $2^+$ 76.5 keV                             & all four spectra      & $-45.1\pm68.3$  (76.5 keV)    & 71.8      & $2.0\times10^{17}$ \\
 \hline
  $^{179}$Hf$~\rightarrow$~$^{175}$Yb       & Fit of sum of         & ~                             &  ~        & ~ \\
 all levels                                 & all four spectra      & $-19.0\pm15.3$ (396.3 keV)    & 10.3      & $2.2\times10^{18}$ \\
  \hline
  $^{180}$Hf$~\rightarrow$~$^{176}$Yb       & Fit of sum spectra    & ~                             &  ~        & ~ \\
  $2^+$ 82.1 keV                            & I Ge7, I Ge10, II Ge7 & $-2.5\pm15.2$  (82.1 keV)     & 22.6      & $1.0\times10^{18}$ \\
 \hline
\end{tabular}
\label{tab:other}
\end{center}
\end{table}

\subsection{Theoretical $T_{1/2}$ estimations}

We calculated theoretical half-life values using 3 different
approaches: (1) semiempirical formulae \cite{Poenaru:1983} based
on the liquid drop model and the description of the $\alpha$ decay
as a very asymmetric fission process; (2) cluster model of
\cite{Buck:1991,Buck:1992}; (3) semiempirical formulae
\cite{Denisov:2015}. These approaches were tested with known
experimental half-lives of near four hundred $\alpha$ emitters and
demonstrated good agreement between calculated and experimental
$T_{1/2}$ values, mainly inside a factor of $2-3$. No change in
spin and parity in parent-to-daughter transition is supposed in
calculations with \cite{Poenaru:1983,Buck:1991,Buck:1992}. So, in
the cases, when such a difference existed and the emitted $\alpha$
particle has non-zero angular momentum $L$, we took into account
the additional hindrance factor, calculated in accordance with
\cite{Heyde:1999} (for the lowest possible $L$ value). Denisov et
al. \cite{Denisov:2015} took non-zero $L$ into account explicitly.

Table~\ref{tab:res} summarizes the experimental half-life limits
(lower bounds) obtained in this work as well as the half-life
values obtained from the theoretical calculations.

\begin{table}[ht]
\caption{Half-life limits on the $\alpha$ decay of Hf isotopes in
comparison with the theoretical predictions calculated here with
approaches \cite{Poenaru:1983,Buck:1991,Buck:1992,Denisov:2015}.
The limits are given at 90\% C.L.}
\begin{center}
\begin{tabular}{|l|l|l|l|l|l|}
\hline
Transition                  & Level of the     & Experimental                                  & \multicolumn{3}{c|}{Theoretical $T_{1/2}$ (a)} \\
\cline{4-6}
~                           & daughter         & $T_{1/2}$ (a)                                & \cite{Poenaru:1983} & \cite{Buck:1991,Buck:1992} & \cite{Denisov:2015} \\
~                           & nucleus (keV)    & ~                                             & ~                   & ~                          & ~ \\
\hline
$^{174}$Hf$~\to$~$^{170}$Yb & $0^+$, g.s.      & $=2.0(4)\times10^{15}$ \cite{Macfarlane:1961} & $7.4\times10^{16}$  & $3.5\times10^{16}$         & $3.5\times10^{16}$ \\
~                           & $2^+$, 84.3      & $\geq 3.3\times10^{15}$                       & $3.0\times10^{18}$  & $1.3\times10^{18}$         & $6.6\times10^{17}$ \\
$^{176}$Hf$~\to$~$^{172}$Yb & $2^+$, 78.7      & $\geq 3.0\times10^{17}$                       & $3.5\times10^{22}$  & $1.3\times10^{22}$         & $4.9\times10^{21}$ \\
$^{177}$Hf$~\to$~$^{173}$Yb & $7/2^-$, 78.6    & $\geq 1.3\times10^{18}$                       & $1.2\times10^{24}$  & $9.1\times10^{21}$         & $3.6\times10^{23}$ \\
$^{178}$Hf$~\to$~$^{174}$Yb & $2^+$, 76.5      & $\geq 2.0\times10^{17}$                       & $8.1\times10^{25}$  & $2.4\times 0^{25}$         & $7.1\times10^{24}$ \\
$^{179}$Hf$~\to$~$^{175}$Yb & $(7/2^-)$, g.s.  & $\geq 2.2\times10^{18}$                      & $4.0\times10^{32}$  & $4.4\times10^{29}$         & $4.7\times10^{31}$ \\
~                           & $(9/2)^+$, 104.5 & $\geq 2.2\times10^{18}$                      & $2.5\times10^{35}$  & $2.0\times10^{32}$         & $2.2\times10^{34}$ \\
$^{180}$Hf$~\to$~$^{176}$Yb & $2^+,$ 82.1      & $\geq 1.0\times10^{18}$                       & $4.1\times10^{50}$  & $4.0\times10^{49}$         & $2.1\times10^{48}$ \\
\hline
\end{tabular}
\label{tab:res}
\end{center}
\end{table}

\section{CONCLUSIONS}

A 179.8 g hafnium sample of natural isotopic abundance was
measured using ultra low-background $\gamma$-ray spectrometry with
the aim to search for $\alpha$ decays of the naturally occurring
hafnium isotopes to the first excited state of the daughter
nuclei. For $^{179}$Hf also $\alpha$ decay to the ground state of
$^{175}$Yb was searched for thanks to the $\beta$-instability of
the daughter nuclide $^{175}$Yb. None of the decays were detected
but for the first time lower bounds were derived for these decays
as reported in Table \ref{tab:res}. One can note that among the
different decays, it is the decay of $^{174}$Hf that has the
experimental limit ($3.3 \times10^{15}$ a) which is closest to the
lowest theoretical estimation ($6.6\times10^{17}$ a). There is a
factor 200 between these two values. One cannot exclude that it is
possible to conceive an experiment that succeeds in improving
sensitivity by two orders of magnitude compared to these
measurements. Solely by using a sample of the same mass but
enriched to 10\% (very expensive of course) one obtains a factor
60. By using HPGe detectors optimized for high resolution and low
background in the 84 keV region (e.g. so-called BEGe-detectors)
one possibly obtains another factor 2. Finally by increasing the
measurement time from 2 months to one year one obtains another
factor of 2.5 and in total one reaches a factor 300.

To go further one can go for a ``source = detector'' approach
where the sample and detector are the same. As possible
candidates, two new crystal scintillators can be named:
Cs$_2$HfCl$_6$ \cite{Burger:2015} and Tl$_2$HfCl$_6$
\cite{Fujimoto:2018}. The scintillating bolometer technique with
simultaneous measurement of the heat and light signals seems to be
the best choice ensuring high efficiency ($\simeq 100\%$), good
energy resolution (on the level of few keV), and possibility to
distinguish between signals induced by $\alpha$ and
$\beta(\gamma$) particles. It would be also interesting to
remeasure $^{174}$Hf $\alpha$ decay to the g.s. of $^{170}$Yb
because $\simeq 20-35$ factor difference between the experimental
$T_{1/2}=2.0(4)\times10^{15}$~a \cite{Macfarlane:1961} and
calculated $T_{1/2}=(3.5-7.4)\times10^{16}$~a
\cite{Poenaru:1983,Buck:1991,Buck:1992,Denisov:2015} is not usual
for even-even nuclei (usually this difference is within a factor
$2-3$).

The limits obtained for other hafnium isotopes $\lim T_{1/2}\sim
10^{17-18}$~a are very far from the theoretical predictions.

We have found the hafnium sample contaminated on the level of tens
mBq/kg by $^{232}$Th, $^{235}$U and $^{238}$U daughters. Deep
purification of hafnium looks a rather complicated task, taking
into account typically high contamination of hafnium by Th and U,
and very high melting and boiling points of Hf (2227  $^{\circ}$C
and 4602  $^{\circ}$C, respectively). The last circumstance
provides certain difficulties to apply the efficient purification
methods, like distillation, crystallization or zone melting. The
observed in the sample cosmogenic $^{175}$Hf and $^{181}$Hf on the
mBq/kg level are rather short-living (the half-lives are $\simeq
70$ d and $\simeq 42$ d, respectively). Therefore the
contamination should not provide a substantial background after
long enough (order of year) Hf cooling deep underground. It should
be stressed, however, that the internal radioactive contamination
of the Hf sample utilized in the present experiment does not
restrict the present experiment sensitivity. The main background
was due to the naturally occurring trace radioactive contamination
of the experimental set-ups. Nevertheless, advanced purification
methods and enabling production of pure isotopically enriched
$^{174}$Hf, without contamination by enriched $^{235}$U, would
improve the background situation further.

\section{ACKNOWLEDGMENTS}

This project received support from the EC-JRC open access project
EUFRAT under Horizon2020. The group from the Institute for Nuclear
Research (Kyiv, Ukraine) was supported in part by the program of
the National Academy of Sciences of Ukraine ``Fundamental research
on high-energy physics and nuclear physics (international
cooperation)''. D.V.K. and O.G.P. were supported in part by the
project ``Investigations of rare nuclear processes'' of the
program of the National Academy of Sciences of Ukraine
``Laboratory of young scientists'' (Grant No. 0118U002328).

\end{document}